# Thermodynamics of Magnets

Thomas R. Lemberger
Dept. of Physics
Ohio State University
Columbus, OH 43210-1106

Thermodynamics of magnetic materials is discussed in practical, lab-oriented terms. In the common experimental configuration in which the external magnetic field comes from a solenoidal coil connected to a power supply, magnetic work is identified unambiguously as the flow of electromagnetic field energy from the power supply into the system *via* the connecting wires. A simple algebraic expression is derived for the "magnetic energy" of microscopic dipoles which interact with the magnetic fields produced by each other, by an external coil, or by a permanent magnet. The discussion delineates the important distinction between induced magnetic moments, which are diamagnetic, and permanent microscopic moments, which are paramagnetic. The practicality of these ideas is illustrated by calculations of the magnetic properties of several idealized magnetic solids *via* minimization of the appropriate free energy.

1. Introduction.

50 years ago, Guggenheim[1] noted and attempted to dispel the confusion which he perceived to surround magnetic work and thermodynamics of magnets. Through the intervening decades many texts[*e.g.*, 2-7] and articles[*e.g.*, 8,9] have appeared with the same intention, yet confusion persists. The goal of the present article is to distill thermodynamics of magnets from these works, gaining clarity at a minor cost in generality, and to provide example applications to model systems, so that the subject could be included in a course on thermodynamics. Most confusion centers on magnetic work, so we will begin with a common laboratory setup in which magnetic work can be identified unambiguously. The next largest source of confusion is magnetic energy, which is usually expressed as a volume integral involving magnetic fields and/or magnetization. We will acknowledge that the important microscopic objects are dipoles, do the necessary integrals analytically, and obtain a simple algebraic expression. Finally, we will calculate the equilibrium properties of a variety of prototypical magnetic materials by minimizing their free energies. MKS units are used throughout. $\mu_0 \equiv 4\pi \times 10^{-7}$ H/m is the permeability of vacuum.

2. The Lab, the System, and Magnetic Work.

Consider an ellipsoidal homogeneous sample which experiences the magnetic field, $\mathbf{B}_e(\mathbf{r})$, produced by a current flowing in an external coil. $\mathbf{B}_e$ written without an explicit argument is the field at the sample, assumed to be uniform over the sample and parallel to $\mathbf{z}$. As shown in Fig. 1, a coax carries current, $i(t)$, to the coil from a power supply. A voltmeter reads the voltage, $V(t)$, across the coax. The $\mathbf{x}$, $\mathbf{y}$, and $\mathbf{z}$ axes coincide with the ellipsoidal axes, $\mathbf{a}$, $\mathbf{b}$, and $\mathbf{c}$, of the sample. If the sample is anisotropic, then the principle axes of its anisotropy coincide with $\mathbf{a}$, $\mathbf{b}$, and $\mathbf{c}$.[10] The sample is in contact with a thermal reservoir at $T_0$ and a pressure reservoir at $P_0$, *e.g.*, a bath of liquid He, whose thermodynamic and magnetic properties are known.

Our thermodynamic system comprises the entropy and the kinetic and potential energy of particles in the sample and a portion of the energy in the magnetic field surrounding the sample. There is flexibility in the choice of magnetic field energy to include because the static field has no entropy.[8] The best choice[2,9] is the *extra* magnetic field energy due to the sample. For purposes of calculating it, (Sec. 4), the closed surface that bounds the system has linear dimensions many times larger than the largest dimension of the coil, and therefore encloses essentially all of the magnetic field energy produced by the current in the circuit as well as the field energy of each of the microscopic magnetic dipoles in the sample. The surface is not so large that it includes the power supply and the voltmeter. The *total* magnetic work done on the system is the electromagnetic energy which flows, *via* Poynting's vector[11], into the system at the place where the coax intersects the surface. The surface is chosen so that the integral of Poynting's vector over the rest of the surface is negligible. We want only the *extra* work done due to the presence of the sample, which corresponds with our choice of magnetic energy.

The extra work done by the power supply in bringing the current from $i$ to $i + di$ in time interval between $t$ and $t + dt$ can be expressed equally well in terms of current and voltage and in terms of magnetic moment and external field. Let $\mathbf{I}(t)$ be the total magnetic moment of the sample. Suppose that between $t$ and $t + dt$, $\mathbf{I}$ changes by $d\mathbf{I}$ and the external field changes by $d\mathbf{B}_e$. The total work done by the power supply is:[1-6] $\int dV\, d[B_e^2(\mathbf{r})/2\mu_0] + \mathbf{B}_e(t) \cdot d\mathbf{I} + i^2 R dt$, where R is the resistance of the circuit. The extra work due to the sample is clearly $\mathbf{B}_e(t) \cdot d\mathbf{I}$. The total work also is $dt$ times the integral of Poynting's vector over the surface of the system, *i.e.*, the piece of surface lying inside the coax. It is elementary to show that the integral equals $V(t)\, i(t) dt$. The result is general, and not restricted to coaxial cables. The extra work done on the system when the sample is present is then: $\Delta V(t)\, i(t) dt$, where $\Delta V(t)$ is the extra voltage.

It is useful to examine how $I_z$ *vs.* $B_e$ is determined from $V(t)$ *vs.* $i(t)$ in the following gedanken experiment. With the sample absent, $\mathbf{B}_e$ is measured at the sample position as a function of $i$. Then $i(t)$ is ramped slowly 100 times from 0 to a maximum value while $V(t)$ is measured. The 100 voltage ramps are averaged to provide an empty-coil baseline, $<V(t)>$ *vs.* $t$. Variations of each empty-coil run about the average measure noise in the apparatus. The sample is introduced with its ellipsoidal **c** axis parallel to $\mathbf{B}_e$. $V(t)$ is measured continuously for 100 more identical current ramps. By Faraday's law applied to the series circuit of voltmeter-coax-coil, the voltmeter voltage, $V(t)$, is $i(t)R$ plus the rate of change of magnetic flux linking the circuit, which is dominated by the flux linking the coil. Subtracting the baseline average, $<V(t)>$, from each of the $V(t)$ ramps taken with the sample present removes the voltage due to the resistance and self-inductance of the coil and leaves only the *extra* voltage, $\Delta V(t)$, due to the *extra* magnetic flux linked into the circuit by the magnetic moment of the sample.

The extra work done between $t'$ and $t' + dt'$ can be expressed as $B_e(t') dI_z$ or as $\Delta V(t') i(t') dt'$. $I_z(t)$ for the j'th ramp can be found by integrating measured quantities:

$$I_{z,j}(t) = I_{z,j}(0) + \int_0^t dt' \frac{\Delta V_j(t') i(t')}{B_e[i(t')]} \tag{1}$$

We assume nonhysteretic samples, so $I_{z,j}(0) = 0$. The equilibrium moment, $<I_z>$ *vs.* $B_e$ at fixed external temperature and pressure is obtained from Eq. (1) with the ensemble average, $<\Delta V(t)>$, replacing $\Delta V_j(t)$. Fluctuations about the average, *i.e.*, $<[I_{z,j}(t) - <I_z(t)>]^2>$, after correction for noise in the apparatus, measure thermal fluctuations in $I_z$.

There are other, more precise ways to measure $I_z$. [See, for example, refs. 12, 13.] If the magnetic field should be produced by a permanent magnet instead of a coil, then we would consider the permanent magnet to be inside the surface which bounds the system so that Poynting's vector is essentially zero over the surface, and no electromagnetic work is ever done. It is still possible for work to be done on the system by the pressure reservoir or by an external agent which moves the sample.

3. Energy, Entropy, and Minimization of Free Energy.

One of the chief utilities of thermodynamics in magnetism is that it permits simple calculations of the equilibrium properties of model systems *via* maximization of total entropy, $S + S_0$, of system plus thermal reservoir, thereby providing an alternative viewpoint to quantum statistical mechanical calculations. We begin by equating the energy gained by the system to the energy lost by reservoirs during an interval $\Delta t$. Let N be the number of moles of formula units of our system, and assume that no chemistry occurs, and that N is constant. Let U be the energy of the system, including the extra magnetic field energy and the kinetic and potential energies of the particles in the sample. Let V be the volume of the sample and $V_0$ be the volume of the pressure reservoir. Assume that the total volume, $V_0 + V$, is fixed. Finally, let $\Delta Q$ represent a small heat transfer from the heat reservoir *into* the system. For a general change in the system in a small time interval, $\Delta t$, conservation of energy requires:

$$\Delta U = \Delta Q + P_0 \Delta V_0 + V(t)i(t)\Delta t$$

$$= \Delta Q - P_0 \Delta V + B_e \Delta I_z. \tag{2}$$

The second equality relies on: $\Delta V_0 = -\Delta V$ and $V(t)i(t)\Delta t = B_e(t)\Delta I_z$.

Now we introduce entropy and its relationship to heat flow and temperature by replacing $\Delta Q$ with $-T_0 \Delta S_0$ in Eq. (2). We presume that the system can exist in many different states that are characterized by an entropy, S, and that might be the equilibrium state of the system under certain externally imposed conditions. Although some of these states are, in fact, never the equilibrium state, they can be reached through fluctuations of the system about its equilibrium state. We do not concern ourselves with the many other, "nonequilibrium" states of the system, which are not characterized by an entropy. The trick is to identify the unconstrained system variables that are free to fluctuate. In chemical thermodynamics, these variables include the concentrations of various molecules in the system, variables that cannot be identified by thermodynamic measurements and must be identified by other means like spectroscopy. The condition of maximum entropy requires that for a system near equilibrium, $d(S + S_0) \leq 0$ for any combination of small changes, "$dX_i$", in its unconstrained system variables, $X_i$. If, for example, we considered a crystalline solid containing permanent microscopic moments distributed on two inequivalent sites, "A" and "B", and one type of induced diamagnetic moment, then we would express $\Delta I_z$ in Eq. (2) as $\Delta I_{A,p,z} + \Delta I_{B,p,z} + \Delta I_{d,z}$. For such a system, maximization of entropy leads in the usual way to:

$$d[U - T_0 S + P_0 V - B_e(I_{A,p,z} + I_{B,p,z} + I_{d,z})] \geq 0, \tag{3}$$

for fixed $T_0$, $P_0$, and $\mathbf{B}_e$. Thus, the "free energy" to be minimized is:

$$G_0 \equiv U - T_0 S + P_0 V - B_e(I_{A,p,z} + I_{B,p,z} + I_{d,z}). \tag{4}$$

A subtlety here is that the net diamagnetic moment, $I_{d,z}$, is not an unconstrained variable. We know from quantum mechanics that the diamagnetic moment of an atom is the ineluctable response of its core and valence electrons to the magnetic field that they feel. Thus, $I_{d,z}$ is a function of the extensive system variables: U, V, N, and the paramagnetic moments, $I_{A,p,z}$ and $I_{B,p,z}$, and of the external field, $B_e$. This quirk confounds efforts to fit diamagnetism into a thermodynamic formalism which begins with the premise that S for potential equilibrium states is as a function of extensive system variables only.[14] Given that the entropy is nonzero only for nonzero temperatures, it is better to start with the premise that S and U for potential equilibrium states are functions of T and extensive system variables V, N, and paramagnetic moments, which we do in the examples below. If we restrict ourselves to equilibrium states only, then S is a well defined function of extensive system parameters, U, V, N, and the total magnetic moment, $I_z$.

4. Magnetic Energy.

We now calculate the "magnetic energy" of permanent and induced magnetic dipole moments. This includes the change in internal energy of each dipole caused by the field that it feels, the self-field energy of each dipole, and the extra field energy due to the overlap of dipole fields with each other and with the applied magnetic field. The integrals can be done analytically because all fields are divergenceless and integrals are effectively over all space. We will calculate the magnetic energy for the two interesting cases where the external field is produced by a coil or by a nearby permanent magnet. In both cases the field is assumed to be uniform over the sample and parallel to $\mathbf{z}$.

The extra energy in the magnetic field is:

$$U_{field} = \int dV \frac{[\mathbf{B}_e(\mathbf{r}) + \mathbf{B}_{dip1}(\mathbf{r}) + \mathbf{B}_{dip2}(\mathbf{r}) + ...]^2 - B_e^2(\mathbf{r})}{2\mu_0}. \tag{5}$$

The total field at $\mathbf{r}$ is $\mathbf{B}_e(\mathbf{r})$ plus the field due to each dipole, $\boldsymbol{\mu}_j$, in the sample. There are 3 types of terms in Eq. (5). The extra field energy due to overlap of external field and dipole fields can be calculated analytically:

$$\mu_0^{-1}\int dV\, \mathbf{B}_e(\mathbf{r}) \cdot [\mathbf{B}_{dip1}(\mathbf{r}) + \mathbf{B}_{dip2}(\mathbf{r}) + \ldots] = \mathbf{B}_e \cdot (\boldsymbol{\mu}_1 + \boldsymbol{\mu}_2 + \ldots). \qquad (6)$$

The field energy due to overlap of the fields of, *e.g.*, dipoles 1 and 2 can be calculated analytically:

$$\mu_0^{-1}\int dV\, \mathbf{B}_{dip1}(\mathbf{r}) \cdot \mathbf{B}_{dip2}(\mathbf{r}) = \mathbf{B}_{dip1}(\mathbf{r}_2) \cdot \boldsymbol{\mu}_2 = \mathbf{B}_{dip2}(\mathbf{r}_1) \cdot \boldsymbol{\mu}_1. \qquad (7)$$

The self-field energy of a dipole is: $\int dV\, B_{dipj}(\mathbf{r})^2/2\mu_0$. There is no need to evaluate the integral. For induced dipoles, this energy is negligible. For permanent dipoles, this energy is constant and can be absorbed into the rest mass energy.

If we can neglect diamagnetism, so that all microscopic dipoles are permanent, then the magnetic energy is particularly simple. The field from a coil changes the internal energy of each dipole by $-\mathbf{B}_e \cdot \boldsymbol{\mu}$, which cancels against the extra field energy in Eq. (6). Dipole 1 changes the internal energy of dipole 2 by: $-\mathbf{B}_{dip1}(\mathbf{r}_2) \cdot \boldsymbol{\mu}_2$, and *vice versa*. Because $\mathbf{B}_{dip1}(\mathbf{r}_2) \cdot \boldsymbol{\mu}_2 = \mathbf{B}_{dip2}(\mathbf{r}_1) \cdot \boldsymbol{\mu}_1$, the net interaction between two permanent dipoles can be written as $-\mathbf{B}_{dip1}(\mathbf{r}_2) \cdot \boldsymbol{\mu}_2$ or as $-\mathbf{B}_{dip2}(\mathbf{r}_1) \cdot \boldsymbol{\mu}_1$. If both dipoles are in the sample, we assign half to each. If one is in the sample and one is in a permanent magnet nearby, we assign all of the energy to the one in the sample. The net magnetic energy of permanent dipole $\boldsymbol{\mu}_j$ is then:

$$U_{j,mag} = -\boldsymbol{\mu}_j \cdot [\mathbf{B}_{loc}(\mathbf{r}_j)/2 + \mathbf{B}_{pm}], \qquad (8)$$

where $\mathbf{B}_{loc}(\mathbf{r}_j)$ is the field at $\mathbf{r}_j$ due to all other permanent moments in the sample and $\mathbf{B}_{pm}(\mathbf{r}_j)$ is the field from dipoles in the permanent magnet. Typically, $\mathbf{B}_{pm}$ is uniform over the sample. If, in addition, all of the dipoles in the sample occupy equivalent sites, the average local field is the same for all, and summing yields a total energy:

$$U_{p,mag} = -(\mathbf{B}_{loc}/2 + \mathbf{B}_{pm}) \cdot \mathbf{I}_p, \qquad (9)$$

where $\mathbf{I}_p$ is the net dipole moment of the sample.

When induced dipoles are included, the situation is a little more involved. The self-field energy of induced dipoles is so small that it is neglected without comment.[*e.g.*, 15-17] We lump it with the internal "magnetic energy" of the atom or molecule. The internal magnetic energy associated with diamagnetism of an atom or molecule is the field-induced change in kinetic and Coulomb potential energy of the electrons, which, in a calculation, traces back to the vector potential in the kinetic energy operator. Classically, it comes from the change in electron orbits due to the electric field that is induced, *via* Faraday's Law, when the external magnetic field changes. The magnetic energy of the dipole, $\Delta E$, including the change, $\Delta E_{int}$, in its internal energy and the field energy, $\boldsymbol{\mu}_d \cdot \mathbf{B}_e$, [Eq. (6)] is proportional to $B_e^2$ for a linear system. The magnetic susceptibility of a single atom or molecule is best defined[18] as:

$$\chi_{d,mol} \equiv \Delta E /(B_e^2/2\mu_0). \tag{10}$$

The magnetic moment is: $\boldsymbol{\mu}_d = \chi_{d,mol}\mathbf{B}_e/\mu_0$, so we can write the magnetic energy as:

$$\Delta E = \boldsymbol{\mu}_d \cdot \mathbf{B}_e/2, \tag{11}$$

and the internal energy as:

$$\Delta E_{int} = -\boldsymbol{\mu}_d \cdot \mathbf{B}_e/2. \tag{12}$$

In general, $\chi_{d,mol}$ is a tensor. For closed-shell atoms, $\chi_{d,mol}$ ranges from about $-1$ to $-50\times 10^{-6}$ cm$^3$/mole.[18] The minus sign in Faraday's Law ensures diamagnetism, $\boldsymbol{\mu}_d \cdot \mathbf{B}_e < 0$. In a solid composed of these molecules, $\chi_{d,mol}$ may be a function of V.

Deduction of the diamagnetic moment of a single molecule from $\Delta E$ *vs.* $B_e$ follows from conservation of energy. When the field is turned on slowly, the extra work done by the power supply must remain in the system because there is no radiation and no other particle to take energy away. The extra work is $_0\!\int^I B_e(I')dI'$, where the total moment, I', is just the single moment, $\boldsymbol{\mu}_d$. If $\boldsymbol{\mu}_d$ is strictly proportional to field, *i.e.*, $\boldsymbol{\mu}_d = \chi_{d,mol}B_e/\mu_0$, then the extra work is $\int\mu_0(\mu_d'/\chi_{d,mol})d\mu_d' = \mu_0\mu_d^2/2\chi_{d,mol} = \chi_{d,mol}B_e^2/2\mu_0$. Thus, we have Eq. (10) relating $\Delta E$ to $\chi_{d,mol}$.

The total field experienced by diamagnetic dipole j includes the net field, $\mathbf{B}_{loc}(\mathbf{r}_j)$, from other dipoles in the sample and the field, $\mathbf{B}_{pm}$, from the dipoles in a permanent magnet, so $\boldsymbol{\mu}_j = \chi_{d,mol}[\mathbf{B}_e + \mathbf{B}_{loc}(\mathbf{r}_j) + \mathbf{B}_{pm}]$. We assign to dipole j the change in its internal energy, $-\boldsymbol{\mu}_j \cdot [\mathbf{B}_e + \mathbf{B}_{loc}(\mathbf{r}_j) + \mathbf{B}_{pm}]/2$, the field energy, $\boldsymbol{\mu}_j \cdot \mathbf{B}_e$, half of the dipole-dipole field energy, $\boldsymbol{\mu}_j \cdot \mathbf{B}_{loc}(\mathbf{r}_j)/2$, involving other dipoles in the sample, and all of the field energy, $\boldsymbol{\mu}_j \cdot \mathbf{B}_{pm}$, involving dipoles in the permanent magnet. Finally, we assign to each diamagnetic moment the change which its field induces in the internal energy of the permanent dipole moments in the permanent magnet, which we write as: $-\boldsymbol{\mu}_j \cdot \mathbf{B}_{pm}$. The net magnetic energy of each induced dipole in the sample is then:

$$U_{d,j} = \tfrac{1}{2}\,\boldsymbol{\mu}_j \cdot (\mathbf{B}_e - \mathbf{B}_{pm}). \tag{13}$$

Summing over all induced dipoles, assumed to be equivalent, the magnetic energy assigned to the total dipole moment, $\mathbf{I}_d$, is:

$$U_d = \tfrac{1}{2}\, \mathbf{I}_d \cdot (\mathbf{B}_e - \mathbf{B}_{pm}). \tag{14}$$

We assign to permanent magnetic dipole, k, the change in its internal energy, $-\boldsymbol{\mu}_k \cdot [\mathbf{B}_e + \mathbf{B}_{loc}(\mathbf{r}_k) + \mathbf{B}_{pm}]$, the interaction field energy, $\boldsymbol{\mu}_k \cdot [\mathbf{B}_e + \mathbf{B}_{loc}(\mathbf{r}_k)/2 + \mathbf{B}_{pm}]$, and the change in internal energy that the field of dipole k induces in the dipoles in the permanent magnet, $-\boldsymbol{\mu}_k \cdot \mathbf{B}_{pm}$, following the same prescription as for induced dipoles. The net magnetic energy of permanent dipole k is then:

$$U_{p,k} = -\boldsymbol{\mu}_k \cdot [\mathbf{B}_{loc}(\mathbf{r}_k)/2 + \mathbf{B}_{pm}]. \tag{15}$$

If all permanent moments occupy equivalent sites, then summing yields their magnetic energy:

$$U_p = -\mathbf{I}_p \cdot (\mathbf{B}_{loc,p}/2 + \mathbf{B}_{pm}), \tag{16}$$

in terms of the average local field, $\mathbf{B}_{loc,p}$, that they feel and their net dipole moment, $\mathbf{I}_p$. It is easy to generalize to two or more inequivalent sites.

The absence of $\mathbf{B}_e$ from Eq. (15) looks suspicious, but it is what permits paramagnetic moments to fluctuate independently of the energy of the system. Moreover, it is consistent with conservation of energy. Consider the energy of a single isolated permanent moment, $\boldsymbol{\mu}$, while the field from a coil is turned on. The foregoing argues that its energy does not change, requiring that the power supply does no extra work. Let us see. When the field is increased slowly, quantum mechanics tells us that $\boldsymbol{\mu}$ precesses ever more rapidly about $\mathbf{B}_e$, but its projection along $\mathbf{B}_e$ doesn't change. Hence, $\Delta I_z = \Delta\mu_z = 0$, and the power supply does no extra work. Now consider what happens when $\mu_z$ changes in fixed field due to a collision with a passing particle. The power supply does work, $B_e \Delta\mu_z$, while the extra magnetic field energy, $B_e \mu_z$, increases by $\Delta(B_e \mu_z) = B_e \Delta\mu_z$, conserving energy. The scattering event conserves energy separately. The internal energy of the moment *decreases* by $\Delta(B_e \mu_z) = B_e \Delta\mu_z$, while the kinetic energy of the scattering particle *increases* by the same amount.

5. Equivalence between a Coil and a Permanent Magnet.

If the external field is produced by a coil, then the free energy, $G_0$, contains a "Legendre" term, $-B_e(\mu_{d,z} + \mu_{p,z})$, for each dipole in the sample. Combining this with the magnetic energies of induced and permanent microscopic dipoles, Eqs. (13) and (15) with $\mathbf{B}_{pm} = 0$, yields a net contribution to $G_0$ of: $-\mu_{d,z} B_e/2 - \mu_{p,z} B_{loc,p}/2 - \mu_{p,z} B_e$. If the field is produced by a permanent magnet, we take the magnet to reside inside of the surface which bounds the system, discussed in Sec. 2. No magnetic work is done on the system, so there is no Legendre term. With $\mathbf{B}_e = 0$, the net magnetic energy contribution to $G_0$ is the same as it is with a coil, but with $\mathbf{B}_{pm}$ in place of $\mathbf{B}_e$. Thus, in statistical mechanics, the probability of a permanent moment having a particular z component, $\mu_{p,z}$, contains a factor, $exp[-\mu_{p,z}B_{ext}]$, where $B_{ext}$ is the external field, without reference to the source of the field. If the local field, $\mathbf{B}_{loc}$, is obtained in a mean-field theory, then the probability contains an additional factor, $exp[-\mu_{p,z}B_{loc}/2]$, not the naively-expected factor, $exp[-\mu_{p,z}B_{loc}]$.

6. Diamagnets.

6.A. Diamagnetic Susceptibility.

Consider a solid composed of N moles of identical diamagnetic atoms at low temperatures. $\mathbf{I}_d$ is calculated from the magnetic susceptibility of each atom and the field that it feels. Let us assume that each atom sits at a site of cubic symmetry, so the average local field due to atoms within a sphere centered on each atom vanishes. The local field due to atoms outside of the sphere is the average field inside a uniformly magnetized ellipsoid, (the full sample), minus the field inside a uniformly magnetized sphere. With demagnetization factors[19] of 1/3 for a sphere and $\eta_z$ for the sample,

$$B_{loc,z}/\mu_0 = [(1 - \eta_z) - (1 - 1/3)] M_z = (1/3 - \eta_z) I_{d,z}/V. \qquad (17)$$

The magnetization, $M_z \equiv I_{d,z}/V$, is:

$$M_z = NN_A\chi_{d,mol}(B_e + B_{loc,z})/V\mu_0. \qquad (18)$$

$N_A$ is Avogadro's number. By using Eq. (18) in Eq. (17), we can solve for $I_{d,z}$ and obtain the susceptibility, $\chi_D \equiv \mu_0 M_z/B_e$:

$$1/\chi_D = 1/\chi_{D,0} - 1/3 + \eta_z, \qquad (19)$$

where the "bare" susceptibility is:

$$\chi_{D,0} \equiv NN_A\chi_{d,mol}/V. \qquad (20)$$

The conventional susceptibility is defined as $M_z/H_{in,z}$, where $\mathbf{H}_{in} \equiv \mathbf{B}_{in}/\mu_0 - \mathbf{M}$. In the present case, $H_{in,z} = B_e/\mu_0 - \eta_z I_{d,z}/V$. Replacing $B_e/\mu_0$ by $H_{in,z} + \eta_z I_{d,z}/V$ in Eq. (18) yields:

$$1/\chi_{D,conv} = 1/\chi_{D,0} - 1/3, \qquad (21)$$

which is independent of sample shape. Finally, the magnetic energy is [Eq.(14)]:

$$U_{mag} = V\chi_D B_e^2/2\mu_0. \tag{22}$$

6.B. Magnetostriction and Thermal Expansion.

For this example, we examine the origins of magnetostriction and thermal expansion. We add to $G_0$ the Helmholtz free energy, $U_{coh} + U_{ph} - T_{ph}S_{ph}$, of the lattice. $U_{coh}$ is the cohesive energy of the solid at $T = 0$, $B_e = 0$, $P = 0$, and $V = Nv_0$, relative to neutral atoms far apart. If $\kappa$ is the compressibility of the unmagnetized lattice at $T = 0$ and $P = 0$, then:

$$U_{coh.} \approx Nu_{coh.}(V = Nv_0) + [Nv_0 - V - V\ln(Nv_0/V)]/\kappa$$

$$\approx Nu_{coh.}(V = Nv_0) + Nv_0(1 - V/Nv_0)^2/2\kappa. \tag{23}$$

Assuming a typical phonon heat capacity[20], $C_{ph}(T_{ph}, V) \approx ANR(6T_{ph}/\theta_D)^3$, where $\theta_D(V)$ is the volume-dependent Debye temperature, R is the gas constant, and $A \approx 1$, integration of $dU_{ph} = C_{ph}(T')dT'$ and $dS_{ph} = C_{ph}(T')dT'/T'$ from 0 to $T_{ph}$ yields:

$$U_{ph} - T_0 S_{ph} = ANR(6T_{ph}/\theta_D)^3 [T_{ph}/4 - T_0/3]. \tag{24}$$

Therefore,

$$G_0 = \chi_D V B_e^2/2\mu_0 + C_{ph}(T_{ph},V) [T_{ph}/4 - T_0/3]$$

$$+ Nu_{coh.}(V = Nv_0) + Nv_0(1 - V/Nv_0)^2/2\kappa + P_0 V. \tag{25}$$

Minimizing $G_0$ *wrt* $T_{ph}$ at fixed V, N, $P_0$, $B_e$ and $T_0$ yields: $\langle T_{ph}\rangle = T_0$, so we neglect subscripts and replace $T_0$ and $T_{ph}$ by T in Eq. (25). Similarly, we drop the subscript on $P_0$, dropping any distinction between the "external pressure" and the "pressure". The second term in $G_0$ becomes: $-C_{ph}(T,V)T/12$. Minimizing $G_0$ *wrt.* V yields the difference between the equilibrium volume, $\langle V\rangle$, at T, P, and $B_e$ and its value, $Nv_0$, at $T = 0$, $P = 0$ and $B_e = 0$:

$$\langle V\rangle/Nv_0 - 1 = \kappa[P + \partial(\chi_D V)/\partial V|_{T,Be} B_e^2/2\mu_0 - \gamma T C_{ph}/4\langle V\rangle]. \tag{26}$$

We have used the approximate result, $\theta_D(V) \propto V^{-\gamma}$, (the Gruneisen parameter, $\gamma \approx 1$ [21]), so that $\partial C_{ph}/\partial V|_T = 3\gamma C_{ph}/V$. If we evaluate $C_{ph}/\langle V\rangle$ and $\partial(\chi_D V)/\partial V|_{T,Be}$ at $\langle V\rangle = Nv_0$, Eq. (26) is an explicit expression for the equilibrium volume in terms of P, T, and $B_e$. We see that magnetostriction and thermal expansion come from the compressibility of the solid and the volume dependencies of the molecular susceptibility, $\chi_{d,mol}$, [Eq.(10)] and the phonon heat capacity, respectively, for this prototypical diamagnetic solid.

Dropping the <> denotation of equilibrium quantities since all quantities are equilibrium quantities in the following, the equilibrium Gibbs free energy is:

$$G(T,P,B_e,N) = (P - \chi_D B_e^2/2\mu_0)V - C_{ph}(T,V)T/12 + Nu_{coh}(P = 0, T = 0)$$

$$+ Nv_0\kappa[P + \partial(\chi_D V)/\partial V|_{T,B_e} B_e^2/2\mu_0 - \gamma T C_{ph}(T,V)/4V]^2/2, \quad (27)$$

where the equilibrium volume, V, is the function of T, P, $B_e$, and N given as <V> in Eq. (26). The equilibrium energy, U, entropy, S, and magnetic moment are known from the preceding calculations. Alternatively, they can be recovered from G. For example, the derivatives of G *wrt* - T, P, and -$B_e$ are S, V, and $I_z$, as expected for $G \equiv U - TS + PV - B_e I_z$, with $dU = TdS - PdV + B_e dI_z$. Thus, G is the usual Gibbs free energy. It is instructive to note that Maxwell relations obtained from G require that, in general, magnetostriction: $\partial V/\partial B_e|_{T,P} = -\partial I_z/\partial P|_{T,B_e} = [\partial I_z/\partial V|_{T,B_e}][-\partial V/\partial P|_{T,B_e}] = V\kappa_T \partial I_z/\partial V|_{T,B_e}$, and thermal expansion: $\partial V/\partial T|_{P,B_e} = -\partial S/\partial P|_{T,B_e} = [\partial S/\partial V|_{T,B_e}][-\partial V/\partial P|_{T,B_e}] = V\kappa_T \partial S/\partial V|_{T,B_e}$, arise from the isothermal compressibility, $\kappa_T \equiv -V^{-1}\partial V/\partial P|_{T,B_e}$, and the volume dependencies of the diamagnetic moment and the entropy of phonons, respectively, in essential agreement with what we found in the specific example described by Eq. (26).

7. Ferromagnets and Antiferromagnets.

Now consider a crystalline solid in which each atom has a permanent magnetic dipole moment, $\mu$. Ignore diamagnetism. Let all N atoms occupy equivalent sites of cubic symmetry. The local field is given by Eq. (17), with $I_{p,z}$ replacing $I_{d,z}$, and the net magnetic energy [Eq. (16)] is:

$$U_p = -I_{p,z} B_{loc,z}/2 = -\mu_0(1/3 - \eta_z)I_{p,z}^2/2V. \quad (28)$$

Nonmagnetic interactions among localized permanent moments are the focus of a great deal of research on magnetism. Pairwise interactions generally dominate and the nonmagnetic energy is approximately quadratic:

$$U_{nonmag} = -NaI_{p,z}^2/2I_0^2. \quad (29)$$

$I_0 = NN_A\mu$ is the total moment when all dipoles are parallel. *a* is the part of the interaction between one atom and its neighbors which depends on the relative orientation of their dipole moments. In general, *a* depends on the overlap of atomic orbitals, hence on V, leading to magnetostriction as in the previous section. Let us assume that entropy of the permanent moments is largest in the unpolarized state and can be represented by a Taylor expansion for small polarizations, $I_{p,z}/I_0 \ll 1$:

$$S_p/Nk_B \approx s_0 - s_2(I_{p,z}/I_0)^2/2 - s_4(I_{p,z}/I_0)^4/12. \tag{30}$$

$s_0$, $s_2$ and $s_4$ are positive constants of order unity. They take the values: $s_0 = \ln 2$, and $s_2 = s_4 = 1$, for weakly-interacting spin-1/2 systems. The fact that the temperature of the subsystem of permanent moments is zero, by its usual definition as the derivative of $S_p$ *wrt* energy with extensive variables V and $I_{p,z}$ fixed, highlights the difficulty in applying the entropy-based formalism to magnetic systems.

To proceed further, let us simplify the algebra by taking V to be fixed. Conceding that $\langle T_{ph} \rangle = T_0$ and replacing both by T, defining the atomic density, $n \equiv N/V$, and saturation magnetization, $M_{sat} \equiv I_0/V$, $G_0$ becomes:

$$G_0/V = -\mu_0(1/3 - \eta_z)M_z^2/2 - naM_z^2/2M_{sat}^2 - nk_BT(s_0 - s_2M_z^2/2M_{sat}^2$$

$$- s_4M_z^4/12M_{sat}^4) - B_eM_z + U_{coh}/V - 18nAk_B(T/\theta_D)^3. \tag{31}$$

If $a > 0$ and the sample is elongated along **z**, ($\eta_z < 1/3$), favoring ferromagnetism along **z**, this has the classic Ginzburg-Landau form for a weakly polarized, $M_z \ll M_{sat}$, homogeneous ferromagnet[4]:

$$G_0/V = \mu_0M_z^2/2\chi_p + \beta\mu_0M_z^4/4 - B_eM_z + \text{nonmagnetic terms}, \tag{32}$$

where:

$$1/\chi_p \equiv \alpha - 1/3 + \eta_z, \tag{33}$$

$$\alpha \equiv n(s_2k_BT - a)/\mu_0M_{sat}^2, \tag{34}$$

and

$$\beta \equiv s_4nRT/3\mu_0M_{sat}^4. \tag{35}$$

Minimization of $G_0/V$ *wrt* $M_z$ at fixed V and T shows that $\chi_p$ is indeed the magnetic susceptibility at high temperature, and that $1/\chi_p$ becomes negative and a spontaneous moment forms when T drops below the Curie temperature, $T_C \equiv [a + (\mu_0M_{sat}^2/n)(1/3 - \eta_z)]/s_2k_B$.

Antiferromagnetism arises when $a < 0$, as follows. Suppose that nonmagnetic interactions beyond nearest neighbor are negligible. Consider N moments of the same size to be of two types, "A" and "B", which reside on interpenetrating cubic lattices. The nearest neighbors of A's are B's and *vice versa*. The local field for both is $B_{loc,z} = \mu_0(1/3 - \eta_z)(M_{A,z} + M_{B,z})$. The nonmagnetic energy is: $U_{nonmag} \approx N|a|I_{A,z}I_{B,z}/2(I_0/2)^2$. The magnetic entropy is the sum of the entropies of the A and B subsystems. Thus,

$$G_0/V = -\mu_0(1/3 - \eta_z)(M_{A,z} + M_{B,z})^2/2 + 2n|a|M_{A,z}M_{B,z}/M_{sat}^2$$

$$-nk_BT\{s_0 - 2s_2[M_{A,z}^2 + M_{B,z}^2]/M_{sat}^2 - 4s_4[M_{A,z}^4 + M_{B,z}^4]/3M_{sat}^4\}$$

$$- B_e(M_{A,z} + M_{B,z}) + \text{nonmagnetic terms.} \qquad (36)$$

Minimize $G_0/V$ *wrt* $M_{A,z}$ and $M_{B,z}$ separately to find equations for $<M_{A,z}>$ and $<M_{B,z}>$:

$$B_e = -\mu_0(1/3 - \eta_z)(<M_{A,z}> + <M_{B,z}>) + 2n|a|<M_{B,z}>/M_{sat}^2 + 4s_2nk_BT<M_{A,z}>/M_{sat}^2$$

$$+ 16s_4nk_BT<M_{A,z}>^3/3M_{sat}^4. \qquad (37)$$

$$B_e = -\mu_0(1/3 - \eta_z)(<M_{A,z}> + <M_{B,z}>) + 2n|a|<M_{A,z}>/M_{sat}^2 + 4s_2nk_BT<M_{B,z}>/M_{sat}^2$$

$$+ 16s_4nk_BT<M_{B,z}>^3/3M_{sat}^4. \qquad (38)$$

At high T and small fields, cubic terms are negligible. There is a symmetric solution,

$$<M_{A,z}> = <M_{B,z}> = B_e/2[n|a|M_{sat}^{-2} + 2s_2nk_BTM_{sat}^{-2} - \mu_0(1/3 - \eta_z)], \qquad (39)$$

and an antisymmetric solution,

$$<M_{A,z}> = -<M_{B,z}> = \pm B_e/[4ns_2k_BTM_{sat}^{-2} - 2n|a|M_{sat}^{-2}], \qquad (40)$$

which diverges at the Neel temperature: $T_N \equiv |a|/2s_2k_B$. The symmetric solution minimizes $G_0/V$ for $T > T_N$, and we find the usual result for the susceptibility, $\chi_{AF} \equiv \mu_0M_z/B_e$:

$$1/\chi_{AF} = 1/\chi_{AF,0} - 1/3 + \eta_z, \qquad (41)$$

where,

$$1/\chi_{AF,0} = (T_N + T)2s_2k_Bn/M_{sat}^2. \qquad (42)$$

For $T < T_N$ and $B_e = 0$, the cubic terms must be included, and $G_0/V$ is minimized by the antisymmetric solution,

$$<M_{A,z}>/M_{sat} = -<M_{B,z}>/M_{sat} = \pm(3s_2T_N/2s_4T)^{1/2}(1 - T/T_N)^{1/2}, \qquad (43)$$

which describes spontaneous antiferromagnetic ordering as T drops below $T_N$.

8. Pauli Susceptibility of Conduction Electrons.

The conduction electrons in a metal make an interesting contrast to antiferromagnetism because the permanent dipoles come in two types, "up" and "down", but they are mobile, and the number of dipoles of each type is not fixed. The physics of a degenerate gas of conduction electrons is discussed in any text on solid state physics. We consider that the magnetic-moment-up and magnetic-moment-down conduction electrons form distinct subsystems, each with its own chemical potential, energy, entropy, temperature, and magnetic moment. They share the same volume, V. The total number of electrons is fixed, so the chemical potentials of "ups" and "downs" are not independent.

The simplest approximation is to take the density of states per unit volume, N(E), to be independent of electron energy, to take V to be fixed, and to consider T = 0. In this case, the total density n of electrons is related to the chemical potentials, $\mu_{up}$ and $\mu_{down}$, by:

$$n = N(0)(\mu_{up} + \mu_{down}). \tag{44}$$

If $\mu_{up}$ should change by $\delta\mu_{up}$, then $\mu_{down}$ would have to change by $-\delta\mu_{up}$ to keep n constant. The net magnetization is the electron moment, $\mu_B$, times the excess density of ups over downs:

$$M_z = N(0)\mu_B(\delta\mu_{up} - \delta\mu_{down}) = N(0)\mu_B 2\delta\mu_{up}. \tag{45}$$

Converting downs to ups increases the kinetic energy density by:

$$U_{kin}/V = N(0)\delta\mu_{up}^2 = M_z^2/4N(0)\mu_B^2. \tag{46}$$

Taking the effective local field to be the average field inside the metal, because the electrons are mobile, $B_{loc,z} = \mu_0(1 - \eta_z)M_z$, the magnetic energy density of the electrons is [Eq. (16)]:

$$U_{mag}/V = -\mu_0(1 - \eta_z)M_z^2/2. \tag{47}$$

Finally, the free energy to be minimized is:

$$G_0/V = [1/2N(0)\mu_B^2 - \mu_0(1 - \eta_z)]M_z^2/2 - B_eM_z + \text{nonmag. terms.} \tag{48}$$

Minimizing $G_0/V$ *wrt* $M_z$ at fixed V and T yields the Pauli susceptibility, $\chi_P \equiv \mu_0 M_z/B_e$:

$$1/\chi_P = 1/2N(0)\mu_0\mu_B^2 - 1 + \eta_z. \tag{49}$$

$1/\chi_P$ is typically about $10^5$, so the local field correction is negligible. The reader is invited to repeat the calculation including the energy dependence of the density of states, the Coulomb interaction between electrons, and nonzero temperature.

9. Conclusion.

The formalism developed in this paper is general enough to cover any practical magnetic material. Extending it to include superconductors is straightforward but too involved to include here. Finally, it is possible to use the same approach to produce a simple formalism for thermodynamics of dielectric and paraelectric materials located between the plates of a parallel-plate capacitor.

*Acknowledgements*. I am grateful to Donald M. Ginsberg for his insightful suggestions for improving the presentation, and to my colleagues at Ohio State, C. Jayaprakash, Bruce R. Patton, David O. Edwards and Charles A. Ebner, for many useful discussions.

REFERENCES AND FOOTNOTES

1. E.A. Guggenheim, Thermodynamics: an Advanced Treatment for Chemists and Physicists, (North-Holland, Amsterdam, 1967), 6th revised edition, pp. 338-356.

2. A.B. Pippard, Elements of Classical Thermodynamics, (Cambridge Univ. Press, Cambridge, 1957), pp. 23-27.

3. L.D. Landau and E.M. Lifshitz, Electrodynamics of Continuous Media, (Pergamon Press, Oxford, 1960), pp. 126-136 and 146-166.

4. H.B. Callen, Thermodynamics and an Introduction to Thermostatistics, (John Wiley and Sons, NY, 1985), 2nd ed., pp. 81-84 and 479-483.

5. F. Reif, Statistical and Thermal Physics, (McGraw-Hill, NY, 1965), pp. 438-451.

6. C.J. Adkins, Equilibrium Thermodynamics, (McGraw-Hill, UK, 1975), pp. 41-43 and 251-253.

7. J.R. Waldram, Thermodynamics, (Cambridge Univ. Press, London, 1997).

8. V. Heine, "The Thermodynamics of Bodies in Static Electromagnetic Fields," Phil. Mag. $\underline{57}$, 546-552 (1956).

9. H.A. Leupold, "Notes on the Thermodynamics of Paramagnets," Am. J. Phys. $\underline{37}$, 1047-1054 (1969).

10. Even Landau and Lifshitz, ref. 3, simplify with this assumption.

11. J.D. Jackson, Classical Electrodynamics, (John Wiley and Sons, NY, 1975), pp. 189-190.

12. H. Zijlstra, Experimental Methods in Magnetism, (North-Holland, Amsterdam, 1967).

13. Experimental Magnetism, ed. by G.M. Kalvius and R.S. Tebble (John Wiley and Sons, NY, 1979).

14. A popular formalism, ref. 4, presumes that S can be expressed as a function of extensive variables: U, V, N, and $I_z$, but not $B_e$. With this presumption, by using the same method used to prove that the bulk compressibility and heat capacity are positive for systems in equilibrium, one can prove that the magnetic susceptibility is positive. Clearly this mistreats diamagnets.

15. J.H. van Vleck, Theory of Electric and Magnetic Susceptibilities, (Oxford Univ. Press, London, 1932), pp. 20-21.

16. E. Merzbacher, Quantum Mechanics, (John Wiley and Sons, NY, 1970), 2nd ed., p. 448, prob. 11.

17. N. Ashcroft and M.D. Mermin, Solid State Physics, (Holt, Rinehart & Winston, NY, 1976), pp. 648-649.

18. The standard treatment, *e.g.*, refs. 15-17, calculates only the internal energy, $\Delta E_{int}$, neglects the field energy, $\boldsymbol{\mu} \cdot \mathbf{B}_e$, and defines the low-field susceptibility as: $\chi_{d,mol} \equiv -\Delta E_{int}/(B_e^2/2\mu_0)$. This is equivalent to the definition in Eq. (10), but it leaves obscure the connection between work done by the power supply and the change in energy of the system.

19. Ref. 3, Eqs. (4.25) - (4.35) on pp. 26-27. Demagnetization factors, $\eta_j$, are defined such that the spatial average of the induction, $\mathbf{B}_{in}(\mathbf{r})$, inside a uniformly magnetized ellipsoid in a uniform external field, $\mathbf{B}_e$, is: $B_{in,z} = B_{e,z} + (1 - \eta_z)\mu_0 M_z$, and so on for the **x** and **y** components of $\mathbf{B}_{in}$. Similarly, $H_{in,z} \equiv B_{in,z}/\mu_0 - M_z = B_{e,z}/\mu_0 - \eta_z M_z$, *etc.* Constraints are: $0 \leq \eta_j \leq 1$, and $\eta_x + \eta_y + \eta_z = 1$. For a sphere, $\eta_x = \eta_y = \eta_z = 1/3$. For a needlelike sample with length $\ell$ much greater than its largest diameter, b: $\eta_z \approx (b^2/\ell^2)[\ln(2\ell/b) - 1] \approx 0$, and $\eta_x = \eta_y = \frac{1}{2}(1 - \eta_z) \approx 1/2$. For a flattened sample with largest thickness, d, much less than its diameter b: $\eta_z \approx 1 - \pi d/2b \approx 1$, and $\eta_x = \eta_y \approx \pi d/4b \approx 0$.

20. Ref. 17, p. 459.

21. Ref. 17, pp. 492-495.

FIGURE CAPTION

Fig. 1. Schematic of the experiment. The surface that bounds the system for purposes of calculating the magnetic field energy would actually be much larger than the coil.

Figure 1. Lemberger. Am. J. Phys. MS#9024.

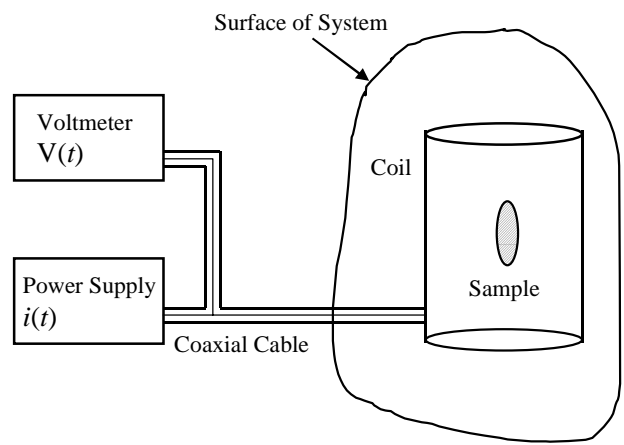